\documentclass[apj,graphicx]{emulateapj}
\usepackage{apjfonts}

\newcommand\mxb{MXB~1659$-$29}
\newcommand\ks{KS~1731$-$260}
\newcommand\chand{{\it Chandra}}
\newcommand\xmm{{\it XMM-Newton}}
\newcommand{\ee}[1]{\ensuremath{\times 10^{#1}}}
\newcommand{\unitspace}{\ensuremath{\,}}
\newcommand{\usp}{\unitspace}
\newcommand{\numberspace}{\ensuremath{\;}}
\newcommand{\nsp}{\numberspace}
\newcommand{\unitstyle}[1]{\ensuremath{\mathrm{#1}}}

\newcommand{\centi}{\unitstyle{c}}
\newcommand{\kilo}{\unitstyle{k}}
\newcommand{\meter}{\unitstyle{m}}
\newcommand{\cm}{\centi\meter}
\newcommand{\gram}{\unitstyle{g}}
\newcommand{\second}{\unitstyle{s}}
\newcommand{\Kelvin}{\unitstyle{K}}
\newcommand{\K}{\Kelvin}
\newcommand{\erg}{\unitstyle{ergs}}
\newcommand{\ergs}{\erg}

\newcommand{\parsec}{\unitstyle{pc}}
\newcommand{\yr}{\unitstyle{yr}}
\newcommand{\Teffinfty}{\ensuremath{T_{\mathrm{eff}}^{\infty}}}
\newcommand{\kTeffinfty}{\ensuremath{k\Teffinfty}}
\newcommand{\Msun}{\ensuremath{M_{\odot}}}

\begin{document}

\title{ 
Cooling of the crust in the neutron star low-mass X-ray binary MXB~1659$-$29}
\shortauthors{Cackett et al.}
\shorttitle{Crustal cooling of MXB~1659$-$29}

\author{Edward~M.~Cackett\altaffilmark{1,2}, 
Rudy~Wijnands\altaffilmark{3},
Jon~M.~Miller\altaffilmark{1},
Edward~F.~Brown\altaffilmark{4},
Nathalie~Degenaar\altaffilmark{3}
}
\email{ecackett@umich.edu}
\altaffiltext{1}{Department of Astronomy, University of Michigan, 500 Church St, Ann Arbor, MI 48109-1042, USA}
\altaffiltext{2}{Dean McLaughlin Postdoctoral Fellow}
\altaffiltext{3}{Astronomical Institute `Anton Pannekoek', University of Amsterdam, Kruislaan 403, 1098 SJ, Amsterdam, the Netherlands}
\altaffiltext{4}{Department of Physics and Astronomy and the Joint Institute for Nuclear Physics, Michigan State University, East Lansing, MI 48824}

\begin{abstract}

In quasi-persistent neutron star transients, long outbursts cause the neutron star crust to be heated out of thermal equilibrium with the rest of the star.  During quiescence, the crust then cools back down.  Such crustal cooling has been observed in two quasi-persistent sources: \ks{} and \mxb{}. Here we present an additional \chand{} observation of \mxb{} in quiescence, which extends the baseline of monitoring to 6.6~yr after the end of the outburst.  This new observation strongly suggests that the crust has thermally relaxed, with the temperature remaining consistent over 1000~days.  Fitting the temperature cooling curve with an exponential plus constant model we determine an $e$-folding timescale of $465\pm25$~days,  with the crust cooling to a constant surface temperature of $kT_{\rm eff}^{\infty} = 54\pm2$~eV (assuming $D=10$ kpc).  From this, we infer a core temperature in the range $(3.5-8.3)\times10^{7}$~K (assuming $D=10$ kpc), with the uncertainty due to the surface composition.  Importantly, we tested two neutron star atmosphere models as well as a blackbody model, and found that the thermal relaxation time of the crust is independent of the chosen model and the assumed distance.
\end{abstract}


\section{Introduction}

Many low-mass X-ray binaries (LMXBs) are transient, spending the majority of their time in a quiescent state with very low levels of accretion and a small fraction of their time in outburst where the mass accretion rate (and hence X-ray luminosity) increases significantly.  For neutron stars, these repeated outbursts affect the star --  the compression of the crust due to accretion of matter induces electron captures, neutron emissions and pycnonuclear reactions \citep{haensel90} in the crust, which in turn heat the core.  Over approximately $10^{4}\textrm{--}10^{5}$~yr a steady-state is reached, in which this deep crustal heating during outburst is balanced by cooling during quiescence \citep{brown98}.

Typically outbursts in these neutron star X-ray transients only last weeks to months; however, in the so-called quasi-persistent transients, outbursts last several years.  In these objects enough heat is imparted to the neutron star crust that it gets heated significantly out of thermal equilibrium with the rest of the star, which is not the case for the majority of normal X-ray transients.  Therefore, in these quasi-persistent sources, once the source returns to quiescence the crust cools down into thermal equilibrium with the core over an appreciable timescale \citep[e.g.][]{rutledge02}.  So far, crustal cooling curves have been observed in two such sources: \mxb{} and \ks{} \citep{wijnands01,wijnands03,wijnands04,cackett06}.  For both objects we obtained several \chand{} and \xmm{} observations over a period of approximately 4 years after the end of long outbursts. \citet{cackett06} found that both sources cooled down rapidly, with the cooling well fit by an exponential decay to a constant level.  This was interpreted as the neutron star crust cooling down into thermal equilibrium with the rest of the star.  By comparison to crustal cooling models of \citet{rutledge02} the cooling curves suggest that the crusts both have a high thermal conductivity, and the cores may require enhanced levels of neutrino emission.  These observations motivated further theoretical study of crustal cooling, and models calculated by \citet{shternin07} for \ks{} also have a best fit with a high thermal conductivity crust, but may not require enhanced neutrino emission in the core. In this paper, we present an additional \chand{} observation of \mxb{}, extending the quiescent monitoring by 2.8 yr to now cover 6.6 yr after the end of the outburst.  In section \ref{sec:results} we detail the data analysis and present the results, while in section \ref{sec:disc} we discuss the cooling curve and implications for the neutron star structure.
 
\section{Analysis and Results} \label{sec:results}

The neutron star low-mass X-ray binary \mxb{} was in outburst for about 2.5~yr, with the source going into quiescence in September 2001.  Since then, the source had been observed 6 times \citep[5 with \chand{}, 1 with \xmm{}; see][]{wijnands03,wijnands04,cackett06}. On 2008 April 27 (MJD 54583.8) a 7th quiescent observation was performed -- we observed \mxb{} for approximately 28~ksec with the \chand{} ACIS-S (ObsID 8984).  The source was at the default aim-point on the S3 chip and was observed using the standard FAINT data mode.  These data were reduced using the latest \chand{} software (CIAO version 4.0) and calibration databases (CALDB version 3.4.3).  No background flares were observed, so all data is used.  A circular region centered on \mxb{} with a radius of 3 arcsec was used to extract the source spectrum.  For the background spectrum we used an annulus centered on the source with an inner radius of  7 arcsec and an outer radius of 22 arcsec.  The source and background spectra were extracted using the  CIAO tool \verb|psextract|, with the correct rmf and arf created using the \verb|mkacisrmf|  and \verb|mkarf|  tools.

\mxb{} is seen to exhibit X-ray eclipses and dips, with an orbital period of $\sim$7.1~hr and an eclipse length of 900~s \citep{cominsky84,cominsky89,oosterbroek01}.  During the eclipses we do not expect to detect the source and in the very first \chand{} observation of \mxb{} in quiescence, the eclipse was clearly seen in the lightcurve \citep{wijnands03}. Using the ephemeris of \citet{oosterbroek01}, we determined that only one eclipse occurred during our observation and we therefore reduced the exposure time in the spectrum by 900~s to account for the fact that we would not receive any counts during the eclipse.  After this correction, the background-subtracted net count rate in the 0.5-10~keV band is $(1.0\pm 0.2) \times10^{-3}$ counts~s$^{-1}$, with the corrected exposure time 26.7~ks.

\subsection{Spectral Fitting}

\begin{deluxetable*}{lccccccc}
\tabletypesize{\scriptsize}
\tablecolumns{8}
\tablewidth{0pc}
\tablecaption{Spectral fitting parameters}
\tablecomments{All uncertainties are 1$\sigma$. The mass and radius are fixed in all fits to 1.4 M$_\odot$ and 10 km respectively.  In the nsatmos fits the normalization is fixed at 1.0. ObsIDs of each observation are indicated at the top, with CXO (\chand{}) and XMM (\xmm{}) denoting the observatory.}
\tablehead{ Parameter & 2688 & 3794  & 0153190101 & 3795  & 5469 & 6337 & 8984 \\
 & (CXO) & (CXO) & (XMM) & (CXO) & (CXO) & (CXO) & (CXO)}
\startdata
MJD & 52197.8 & 52563.2 & 52712.2 & 52768.9 & 53560.0 & 53576.7 & 54583.8 \\
\hline
 & \multicolumn{7}{c}{NSA, $D=10$ kpc} \\
\hline
N$_{\rm H}$ ($10^{21}$ cm$^{-2}$) & \multicolumn{7}{c}{$2.0\pm0.2$} \\
$kT_{\rm eff}^{\infty}$ (eV) & $121\pm1$ & $85\pm1$ & $77\pm1$ & $73\pm1$ & $58\pm2$ & $54\pm3$ & $56\pm2$\\ 
$F_{\rm bol}$\tablenotemark{a} (10$^{-14}$ erg cm$^{-2}$ s$^{-1}$) & 
$41\pm2$ & $10\pm1$ & $6.6\pm0.3$ & $5.1\pm0.3$ & $2.0\pm0.3$ & $1.5\pm0.3$ & $1.8\pm0.3$\\
\hline
 & \multicolumn{7}{c}{NSA, $D = 5$ kpc} \\
\hline
N$_{\rm H}$ ($10^{21}$ cm$^{-2}$) & \multicolumn{7}{c}{$2.9\pm 0.1$} \\
$kT_{\rm eff}^{\infty}$ (eV) & $95\pm1$ & $69\pm1$ & $63\pm1$ & $59\pm1$ & $48\pm1$ & $45\pm2$ & $47\pm2$\\ 
$F_{\rm bol}$ (10$^{-14}$ erg cm$^{-2}$ s$^{-1}$) & 
$59\pm2$ & $16\pm1$ & $11\pm1$ & $9.0\pm0.6$ & $3.7\pm0.4$ & $2.9\pm0.5$ & $3.5\pm0.5$  \\
\hline
 & \multicolumn{7}{c}{NSA, $D = 13$ kpc} \\
\hline
N$_{\rm H}$ ($10^{21}$ cm$^{-2}$) & \multicolumn{7}{c}{$1.6\pm 0.1$} \\
$kT_{\rm eff}^{\infty}$ (eV) & $133\pm1$ & $93\pm1$ & $84\pm1$ & $79\pm2$ & $62\pm2$ & $57\pm3$ & $61\pm2$\\ 
$F_{\rm bol}$ (10$^{-14}$ erg cm$^{-2}$ s$^{-1}$) & 
$34\pm1$ & $8.1\pm0.5$ & $5.4\pm0.2$ & $4.1\pm0.3$ & $1.6\pm0.2$ & $1.2\pm0.2$ & $1.5\pm0.2$\\
\hline
 & \multicolumn{7}{c}{NSATMOS, $D = 10$ kpc} \\
\hline
N$_{\rm H}$ ($10^{21}$ cm$^{-2}$) & \multicolumn{7}{c}{$2.0\pm0.1$} \\
$kT_{\rm eff}^{\infty}$ (eV) & $121\pm1$ & $86\pm1$ & $78\pm1$ & $73\pm1$ & $57\pm2$ & $54\pm3$ & $57\pm2$ \\ 
$F_{\rm bol}$ (10$^{-14}$ erg cm$^{-2}$ s$^{-1}$) &
$39\pm1$ & $9.8\pm0.5$ & $6.6\pm0.2$ & $5.1\pm0.4$ & $2.0\pm0.3$ & $1.5\pm0.3$ & $1.9\pm0.3$ \\
\hline
 & \multicolumn{7}{c}{BBODYRAD} \\
\hline
N$_{\rm H}$ ($10^{21}$ cm$^{-2}$) & \multicolumn{7}{c}{$1.6\pm0.2$} \\
Normalization\tablenotemark{b}  & \multicolumn{7}{c}{$3.3^{+0.9}_{-0.5}$}\\
$kT_{\rm eff}^{\infty}$ (eV) & $300\pm11$ & $213\pm6$ & $193\pm5$ & $184\pm5$ & $148\pm5$ & $140\pm7$ & $147\pm5$ \\ 
$F_{\rm bol}$ (10$^{-14}$ erg cm$^{-2}$ s$^{-1}$) & 
$29\pm4$ & $7.2\pm0.9$ & $4.9\pm0.5$ & $4.0\pm0.5$ & $1.7\pm0.2$ & $1.3\pm0.3$ & $1.6\pm0.2$
\enddata
\label{tab:specfits}
\tablenotetext{a}{Bolometric flux calculated from the model over the 0.01 - 100 keV range}
\tablenotetext{b}{Normalization for the blackbody model is ($R$/D10)$^2$. $R$ is the emitting radius (km) and D10 is ($D$/10 kpc)}
\end{deluxetable*}

In this analysis we fit this latest \chand{} spectrum simultaneously with the previous 5 \chand{} and 1 \xmm{} observations of \mxb{} in quiescence \citep[we do not create new spectra of these previous observations here; see][]{cackett06}.  For spectral fitting we used \verb|xspec| \citep[version 11,][]{arnaud96}.  The spectra were left unbinned due to the low number of counts and the W-statistic \citep{wachter79} was used in all fits.

Neutron star atmosphere spectra deviate from that of a simple blackbody as they have a slightly harder tail due to the strong frequency dependence of opacity for free-free absorption \citep{zavlin96}.  This difference means that when a blackbody model is fitted to the X-ray spectra of quiescent neutron stars a significantly higher surface temperature and an unrealistically small emitting radius are determined \citep[e.g.,][]{rutledge99}.  The spectra were therefore fitted with an absorbed neutron star atmosphere model.  There are currently a variety of neutron star atmosphere models available for the pure hydrogen, low-magnetic field case that is relevant here.  In \citet{cackett06} we used the \verb|nsa| model \citep{pavlov91,zavlin96}, which we also use here.  In order to test the model dependence of the cooling curves we chose to also fit the \verb|nsatmos| model \citep{heinke06}.  Finally, for completeness, we also fit a simple blackbody model (\verb|bbodyrad|).

In the neutron star atmosphere fits, we fixed the mass and radius at the canonical values of 1.4 M$_\odot$ and 10 km as the data is not of high enough quality to independently determine these parameters.  The normalization in the \verb|nsa| model is defined as $1/D^2$, where $D$ is the distance to the object in pc.  In the \verb|nsatmos| model there is also a distance parameter, as well as a separate normalization, $K$, which corresponds to the fraction of the neutron star surface that is emitting.  We fix $K = 1$ in all our fits with \verb|nsatmos|.  In \citet{cackett06} we fixed $D=10$ kpc to reduce the number of free parameters in the fit given the low number of counts in the majority of the spectra.  If the distance is left as a free parameter the uncertainty in the parameters are dominated by the uncertainty in the distance.  The distance to \mxb{} is estimated to be 10-13 kpc \citep{muno01,oosterbroek01} from using type-I X-ray bursts.    Here we investigated the dependence of the cooling timescale on the assumed distance by performing spectral fits with $D$ = 5, 10 and 13 kpc. \citet{wijnands04} found that when assuming 3 different distances to \mxb{} there was little effect on the $e$-folding timescale.

For the interstellar photoelectric absorption we used the \verb|phabs| model, and this parameter is tied between all the observations.  Therefore in the spectral fits the free parameters are the absorbing column and the effective temperature. Finally, for the one \xmm{} observation we tie the parameters between the MOS1, MOS2 and pn spectra. Results of the spectral fitting are given in Table \ref{tab:specfits}. Note that we quote $kT_{\rm eff}^{\infty}$, the effective surface temperature for an observer at infinity.  The bolometric flux was estimated by extrapolating the unabsorbed model over the 0.01-100 keV range.  All uncertainties quoted and plotted in this paper are 1-$\sigma$.  We note that in \citet{cackett06} the uncertainties quoted on the spectral fits and plotted on the figures are at the 90\% level of confidence not 1-$\sigma$ as is incorrectly written in the text.

\begin{figure}
\centering
\includegraphics[angle=270]{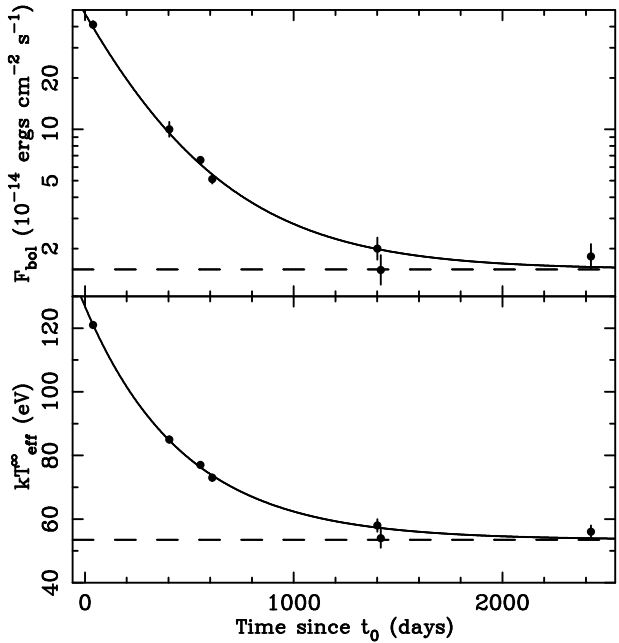}
\caption{Flux (top) and temperature (bottom) cooling curves for \mxb{}. The best-fitting model to the temperatures comprising of an exponential decay to a constant level is shown (solid line), where the constant offset is shown with a dashed line.  The flux cooling curve is then calculated from the fit to the temperatures.  The time t$_0$ = 52159.5 (MJD) is the last day when the source was seen to be active.  Error bars are 1$\sigma$. Data points are from the nsa fits with the distance fixed at 10 kpc.}
\label{fig:cooling}
\end{figure}

All the spectral fits show that the temperature and flux of this new observation has remained consistent with the last two observations performed approximately 1000 days previously, regardless of the model used, or distance assumed, in the fitting.  The temperature and flux determined from the \verb|nsa| fits with $D=10$ kpc is shown in Figure \ref{fig:cooling}. We found that there is very good agreement between the effective temperature  determined by the \verb|nsa| and \verb|nsatmos| fits. Additionally, we found that the temperatures and fluxes from the neutron star atmosphere fits are just shifted up or down in a systematic way depending on the assumed distance. The temperatures determined from the \verb|bbodyrad| fits are significantly higher than the neutron star atmosphere temperatures.  Moreover, if a distance of 10 kpc is assumed, the normalization implies an emitting radius of $\sim$2 km, much smaller than is realistic if the entire surface is emitting. Both high temperatures and small emitting radii are normally found when fitting a blackbody model to quiescent neutron stars \citep[e.g.,][]{rutledge99}.

\subsection{Cooling curves}

The results of the spectral fitting (Table \ref{tab:specfits}) show that after the initial cooling the temperature is consistent with being constant over the last 1000 days.  In \citet{cackett06} we found that the cooling curves for both \mxb{} and \ks{} could be well fit by an exponential that decays to a constant level, interpreted as the neutron star crust cooling back into thermal equilibrium with the core.  With the addition of the new observation of \mxb{} we fitted the cooling curve again with the same model of the form $y(t) = a \exp[-(t-t_0)/b] + c$.  We set $t_0$ to midday on the last day that the source was observed to be active, MJD 52159.5.  This model was fit to the temperatures determined by the \verb|nsa| spectral fits (for all assumed distances) as well as to the results of the blackbody fits, allowing us to test the model dependence of the cooling curves.  We do not fit curves to the \verb|nsatmos| results as they so closely match the \verb|nsa| results. We only fit the cooling curve to the temperatures, as from this fit one can directly calculate the flux via $F = \sigma T_{\mathrm{eff}}^{\infty 4} (R_a/D)^2$, where $R_a$ is the apparent radius for an observer at infinity.

The results of these exponential decay plus constant fits are given in Table \ref{tab:cooling}, and we show the temperature and flux cooling curves for the \verb|nsa| fits (with $D=10$ kpc) in Figure~\ref{fig:cooling}.  There is excellent agreement between the $e$-folding timescales, $b$, when using results from different spectral models and assuming different distances, demonstrating that the cooling timescale is robust. For comparison, the cooling curve values for \ks\ values are $a = 40\pm3$ eV, $b=305\pm47$ days, $c=70.2\pm1.2$ eV fitting the temperatures, assuming $D=7$ kpc.  These are updated from \citet{cackett06} using the correct 1-$\sigma$ uncertainties, and rectifying a minor error in the fitting code.  Note, however, that these values remain consistent with those quoted in \citet{cackett06}.

The exponential plus constant model fits the data well with the reduced $\chi^2$ values all close to 1, and the $\chi^2$ probabilities, $P_\chi$, all close to 0.5, as expected for a good fit.  However, the mathematical solution for the flux from a cooling thin layer is a (possibly broken) power-law \citep{eichler89,piro05}, not an exponential.  We therefore also attempt to fit a single power-law to the data of the form $y(t) = a (t-t_0)^b$, and find that such a model does not fit the data well.  However, inspection of the cooling curve indicates that the middle section of the curve does appear to follow a power-law.  Fitting a power-law to just these middle data points (excluding the first and last observation) achieves a good fit.  For fits to the temperatures from the \verb|nsa| results with $D=10$ kpc, we find the power-law index $b = -0.33\pm0.02$, $\chi^2_\nu = 0.9$ and $P_\chi = 0.46$.  The effective temperature from the last observation is a $4.5\sigma$ deviation from an extrapolation of this best-fitting power-law.  This again strongly indicates that the temperature is now remaining constant, and that the crust is thermally relaxed.

\begin{deluxetable}{lcccc}
\tablecolumns{5}
\tablewidth{0pc}
\tablecaption{Cooling curve parameters from fits to the temperatures}
\tablecomments{The model fit to the cooling curves is of the form $y(t) = a \exp[-(t-t_0)/b] + c$, where $t_0 = 52159.5$. All uncertainties are 1$\sigma$.}
\tablehead{ Parameter & nsa & nsa & nsa & bbodyrad\\
 & $D = 10$ kpc & $D = 5$ kpc & $D = 13$ kpc &  }
\startdata
Normalization, a (eV) & $73\pm2$ & $55\pm1$ & $81\pm2$ & $175\pm12$\\
$e$-folding time, b (days) & $465\pm25$ & $483\pm30$ & $473\pm23$ & $435\pm44$\\
Constant, c (eV) & $54\pm2$ & $45\pm1$ & $58\pm2$ & $142\pm4$\\
$\chi^2_\nu$ & 0.8 & 1.5 & 1.4 & 0.7 \\ 
$P_\chi$ & 0.52 & 0.18 & 0.21 & 0.61 
\enddata
\label{tab:cooling}
\end{deluxetable}

\section{Discussion} \label{sec:disc}

We presented a new observation of the quasi-persistent neutron star X-ray transient \mxb{} in quiescence, extending the quiescent monitoring to 6.6~yr.  Results from the first 6 observations showed that the source had cooled rapidly and indicated that the neutron star crust may have returned to thermal equilibrium with the core.  This new observation shows that the neutron star temperature and flux remained consistent with the previous two \chand{} observations performed approximately 1000 days before.  The model dependence of the thermal relaxation timescale was investigated with 2 different neutron star atmosphere models as well as a simple blackbody model.  Moreover, we assumed 3 different distances to \mxb.  The $e$-folding timescales of the cooling curves from all the spectral fits are consistent with each other, demonstrating the robustness of the measurement. The results are consistent with fits to the first 6 observations \citep{cackett06}.

With the crust thermally relaxed, we can compute the core temperature (here we assume $D=10\nsp\kilo\parsec$).  We integrate the thermal structure equation in the neutron star envelope, following the calculation in \citet{brown02}.  The inferred core temperature is relatively insensitive to the mass of the neutron star (the proper $T_{\mathrm{eff}}$ increases with redshift for a fixed \Teffinfty, but the surface layer becomes thinner with increasing $g$ and reduces the rise in temperature in the  envelope).  There is a significant uncertainty resulting from the depth of the light element layer, however \citep{brown02}. We find the inferred core temperature to range from $3.5\ee{7}\nsp\K$, for a deep He layer (column of $10^{8}\nsp\gram\usp\cm^{-2}$) overlaying a pure Fe layer to $8.3\ee{7}\nsp\K$ for an shallow He layer (column of $10^{4}\nsp\gram\usp\cm^{-2}$) overlaying a layer of heavy  rp-process ashes.  We estimate the time-averaged mass accretion rate for the system to be $7\ee{-11}\nsp\Msun\usp\yr^{-1}$, based on the known outburst behavior of 2 outbursts lasting approximately 2.5 yr, with a quiescent period of 21 yr. We estimated the fraction of Eddington luminosity by calculating the ratio of the average persistent outburst flux to the peak type-I X-ray burst flux \citep[taken from][]{wijnands02}. From this, we estimated the time-averaged crust nuclear heating, assuming the heat deposited in the crust is 1.5 MeV per nucleon \citep[see][]{brown98,rutledge02b}, to be $\sim6\ee{33}\nsp\ergs\usp\second^{-1}$, for a distance of $10\nsp\kilo\parsec$.

As noted previously \citep{cackett06,heinke07}, even for the highest core temperature compatible with $T_{\mathrm{eff}}^{\infty}$, the neutrino luminosity resulting from modified Urca cooling \citep[for a review, see][]{yakovlev04} would still be a factor of $\approx 30$ less than the time-averaged crust nuclear heating. As a check on whether there is a need for enhanced cooling, we also computed the neutrino cooling according to the ``minimal cooling model'' \citep{page04}, which includes the pair breaking and formation (PBF) neutrino emissivity, but with the $^{1}S_{0}$ channel suppressed following \citet{steiner08}. In this case the PBF neutrino luminosity from neutrons in the $^{3}P_{2}$ state is still sufficient to balance the time-averaged nuclear heating, if the core temperature is in the upper half of the range given above. Given the uncertainty in the depth of the light element envelope and the superfluid critical temperatures, we cannot exclude that the neutrino emission is solely due to standard cooling processes.

The exponential plus constant cooling curve fits provide a measure of the thermal relaxation time of the crust.  This relaxation time depends on the crust composition and lattice structure \citep{rutledge02,shternin07}, on the crust thickness and hence surface gravity of the neutron star \citep{lattimer94}, and on the distribution of heat sources (\citealt{shternin07,horowitz08a}; Brown \& Cumming, in preparation). \citet{shternin07} showed that the cooling timescale in \ks\ was best fit by having a high thermal conductivity in the crust, as if it were composed of a locally pure lattice.  This matches molecular dynamics simulations \citep{horowitz07, horowitz08b}, which find that the dense crust plasma does indeed freeze into an ordered lattice with a high thermal conductivity. As in \ks, our fits to the cooling of \mxb\ are again consistent with such an ordered, low-impurity crust. \citet{shternin07} noted that the crust may not have completely thermally relaxed; we find a single power-law decay also fits the cooling curve for \ks{} well, with a power-law index = $-0.12\pm0.01$ and $\chi^2_\nu = 0.2$ when fitting to the temperatures. Further observations are required to determine whether \ks{} has continued to cool following a power-law decay or if it has reached a constant \kTeffinfty\ indicative of a thermally relaxed crust.  \citet{shternin07} note that the cooling of \ks\ can be fit without invoking enhanced neutrino emission in the core.  If \mxb\ does have a higher core neutrino emission than \ks\ then this may imply that the neutron star in \mxb\ is somewhat more massive than the one in \ks\ since most equations of state allow higher levels of neutrino emission with increasing central density.  We note, however, that this conclusion assumes that the time-averaged mass accretion rates in both these objects have remained in a steady state.

\acknowledgements
\vspace{-0.25cm}
JMM gratefully acknowledges support from {\it Chandra}.  EFB is supported by Chandra Award no.~TM7-8003X issued by the Chandra X-ray Observatory Center, and by a ATFP grant NNX08AG76G.

\bibliographystyle{apj}
\bibliography{apj-jour,mxb1659}

\end{document}